\documentclass[showpacs,preprintnumbers,amsmath,amssymb, prb, twocolumn]{revtex4-1}

\usepackage{stmaryrd}
\usepackage{txfonts}
\usepackage{amssymb}
\usepackage{mathrsfs}
\usepackage{graphicx}
\usepackage{dcolumn}
\usepackage{bm}
\usepackage{epsfig}
\usepackage{xcolor}
\usepackage{appendix}

\begin{document}

\title{Supersymmetry in Closed Chains of Coupled Majorana Modes}

\author{Zhao Huang\(^{1}\), Shinji Shimasaki\(^{1}\) and Muneto Nitta\(^{1,2}\)}

\affiliation{\(^{1}\)Research and Education Center for Natural Sciences, Keio University, 4-1-1 Hiyoshi, Kanagawa 223-8521, Japan \\
\(^{2}\)Department of Physics,  Keio University, 4-1-1 Hiyoshi, Kanagawa 223-8521, Japan}


\begin{abstract}
We consider a closed chain of even number of Majorana zero modes with nearest-neighbour couplings which are different site by site generically, thus no any crystal symmetry. Instead, we demonstrate the possibility of an emergent supersymmetry (SUSY), which is accompanied by gapless Fermionic excitations. In particular, the condition can be easily satisfied by tuning only one coupling, regardless of how many other couplings are there. Such a system can be realized by four Majorana modes on two parallel Majorana nanowires with their ends connected by Josephson junctions and bodies connected by an external superconducting ring. By tuning the Josephson couplings with a magnetic flux $\Phi$ through the ring, we get the gapless excitations at $\Phi_{SUSY}=\pm f\Phi_0$ with $\Phi_0= hc/2e$, which is signaled by a zero-bias conductance peak in tunneling conductance. We find this $f$ generally a fractional number and oscillating with increasing Zeeman fields that parallel to the nanowires, which provide a unique experimental signature for the existence of Majorana modes. 
\end{abstract}

\date{\today}

\maketitle

\emph{Introduction.---}
The interplay between particle and condensed matter physics has proved remarkable fertile for the development of modern physics \cite{Wilczek16}. Recently, the longed for Majorana fermion finds its stage in condensed matter physics as a collective excitation \cite{Volovik99, Read00, Kitaev01, Wilczek09}. Majorana fermion is a fermion that is its own antiparticle and described by a real solution of the Dirac equation. In a group of materials called topological superconductors which have spin-triplet Cooper pairing, there are gapless excitations that are mixture of electrons and holes with equal amplitude and spin direction, and thus can be regarded as Majorana fermionic modes. Unpaired Majorana modes can stay at well separated topological defects and each of the modes is immune to the local disturbance due to topological protection, which provide a promising platform for decoherence-free quantum computation \cite{Kitaev01}. Because spin-triplet superconductors are rare in nature, it is convenient to construct the effective Hamiltonian through heterostructures, for example with spin-orbital coupling (SOC), Zeeman field and superconductivity combined \cite{Liang08, Sato09,Sau10,Lutchyn10, Oreg10}, where phenomena that can be explained by Majorana modes have been observed in many experiments \cite{Mourik12, Deng12, Rokhinson12, Nadj14, Sun16, Albrecht16, Deng16, Wiedenmann16, He16}. 

Meanwhile supersymmetry (SUSY) is a symmetry that relates bosons and fermions, and extends the Standard model by finding a brother of every known elementary particles with a difference of half spin \cite{Wess74,Wess92,Fayet75, Fayet76}. Although SUSY was initially proposed to solve the hierarchy problem in particle physics, it has later been proposed in many non-relativistic condensed matter systems such as interacting spin systems, cold atoms  and topological matters \cite{Friedan84,Forster89,Fendley03, Feiguin07, Lee07, Qi09, Yu08,Yu10,Grover14, Ponte14, Huijse15,Rahmani15, Jian15, Krempa16, Hsieh16, Jian17, Sagi17, Li17}.  In particular, SUSY in quantum mechanics appears in time-reversal-invariant topological superconductors and Majorana models with translational symmetry, in which the time-reversal and translational operator changes the fermion parity, thus playing the role of a supercharge \cite{Qi09, Hsieh16}.

\begin{figure}[t]
\psfig{figure=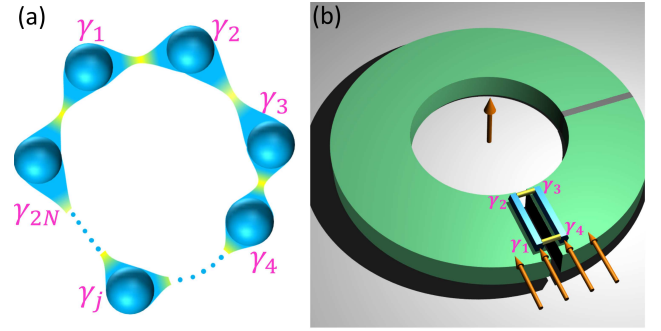,width=8.4cm} \caption{\label{f1} (a) Closed chain of $2N$ Majorana modes with nearest-neighbour couplings without requiring any crystal symmetries. (b) Schematic figure of setup to realize supersymmetry. There are two nanowires (blue) with their ends connected by Josephson junctions (yellow) and bodies connected by an external superconducting ring. The phase shifts across the junctions are controlled by the magnetic fluxes penetrating through the ring. The Zeeman fields parallel to the nanowires are to induce the Majorana modes noted as $\gamma_1, \gamma_2, \gamma_3$ and $\gamma_4$. There is a reference junction (gray) to suppress the phase fluctuation.
}
\end{figure}

In this work we show an experimentally accessible SUSY in a closed chain of coupled Majorana modes without any crystal symmetries, which is different from previously studied translational invariant systems \cite{Hsieh16}. Specifically, we consider even number of Majorana modes with nearest-neighbour couplings as shown in Fig.~\ref{f1}(a). Different from an open chain where the couplings inevitably split the zero-energy levels, we can obtain a nonlocal zero-energy Dirac fermion, resulting in double degeneracy between states of opposite fermion parities at all energy levels, which can be interpreted as a SUSY in quantum mechanics.

We find that despite of the large number of couplings, the SUSY can be reached by tuning only one coupling, which is convenient for experimental realization. The signature of SUSY is a zero-bias peak in tunneling conductance. We design a setup with two parallel Majorana nanowires with their ends linked by Josephson junctions, thus obtaining a closed chain of four Majorana modes with nearest-neighbour couplings. By putting this setup as a part of a superconducting quantum interference device (SQUID) as shown in Fig.~\ref{f1}(b), we can use the magnetic flux $\Phi$ to tune the Josephson couplings between Majorana modes on different nanowires.  In this way, we reach the SUSY at $\Phi_{SUSY}=\pm f\Phi_0$ with $f$ a fractional number in general. In particular, this $f$ oscillates with the Zeeman field that induces the topological superconductivity, which is related to the oscillation of energy splitting caused by hybridization of Majorana modes on a single nanowire \cite{Sarma12}. This fractional number $f$ and its oscillation should be observable in experiments, which provide an indirect demonstration of existence of Majorana modes.

\emph{Supersymmetric closed chain.---} We show the closed chain in Fig.~\ref{f1}(a) where each Majorana mode $\gamma_j$ couples to its nearest neighbours with arbitrary strength. We consider even number of Majorana modes because every operator of Dirac fermion is expressed in terms of two Majorana operators, which makes the even number a natural case. The effective Hamiltonian is given by 
\begin{eqnarray}\label{Hgamma}
H=i\sum\limits_{j=1}^{2N}t_j\gamma_j\gamma_{j+1}=\frac{i}{2}\Gamma^T A \Gamma,
\end{eqnarray}
where $t_j$ is the coupling strength, $\gamma_{2N+1}=\gamma_1$, $\Gamma=(\gamma_1, \gamma_2, .\ .\ .\ , \gamma_{2N})^T$ and $A$ is the corresponding coupling matrix. We do not require any crystal symmetries such as translational, reflection or inversion symmetry for the Hamiltonian. Therefore, generically the Hamiltonian cannot be solved analytically, but an important question is whether exact solutions for low-energy excitations are available in some special occasions. By obtaining the determinant of the coupling matrix $\textrm{Det}(A)=(t_1t_3\cdots t_{2N-1}-t_2t_4\cdots t_{2N})^2$,
it is straightforward to find the existence of zero eigenvalues at the condition
\begin{eqnarray}\label{condition}
\prod_{j=1}^N t_{2j-1}=\prod_{j=1}^N t_{2j},
\end{eqnarray}
which can be easily reached by tuning only one coupling. The open chain indicates only one coupling as zero, which by no means satisfies the above condition and thus no gapless excitation is available.

There are at least two orthogonal zero-energy eigenstates due to the particle-hole symmetry and they are written as
\begin{equation}
\gamma'=|X_1|^{-1}X_1\Gamma, \ \ \ \gamma''=|X_2|^{-1}X_2\Gamma
\end{equation} 
with 
$X_1=\left(1,0,{t_1}/{t_2},0,{t_1t_3}/{t_2t_4},\cdots,{\prod_{j=1}^{N-1}t_{2j-1}}/{\prod_{j=1}^{N-1}t_{2j}},0\right)$ and 
$X_2=\left(0,1,0,{t_2}/{t_3},0,{t_2t_4}/{t_3t_5},\cdots,{\prod_{j=1}^{N-1}t_{2j}}/{\prod_{j=1}^{N-1}t_{2j+1}}\right)$. Here $\gamma'$ and $\gamma''$ are two nonlocal Majorana zero modes with wavefunctions on the whole ring, and combine into a nonlocal gapless Dirac fermion $c=(\gamma'+i\gamma'')/2$.

Now we show all energy levels are at least doubly degenerate. Here we notice that the energy level here means the eigenenergy in many-particle space, not the single-particle excitation energy. We first define the fermion parity 
operator $P=(-i)^N\prod_{j=1}^{2N} \gamma_j$, for which we have $[P, H]=0$ and 
\begin{equation}\label{reverseP}
\gamma'P\gamma'=\gamma''P\gamma''=-P.
\end{equation}
Given $[\gamma',H]=[\gamma'',H]=0$, at all energy levels there are two degenerate states $|\varphi\rangle$ and $\gamma'|\varphi\rangle$ which have opposite fermion parity due to Eq.~(\ref{reverseP}). It is obvious that the degeneracy comes from adding or eliminating one zero-energy Dirac fermion since $\gamma'=c+c^\dagger$, which does not change the total energy but reverses the parity. 

This degeneracy can be interpreted as a SUSY in quantum mechanics. By adding a constant to the Hamiltonian to make all energy levels positive, we can find two fermionic operators
\begin{eqnarray}
Q_1=\gamma'\sqrt{H}, \ \ \ Q_2=\gamma''\sqrt{H},
\end{eqnarray}
which satisfy the algebra
\begin{eqnarray}
\{P,Q_i\}=0,\ \ \{Q_i,Q_j\}=2\delta_{ij}H
\end{eqnarray}
with $i,j\in\{1,2\}$. Therefore, our Hamiltonian exhibits an ${\cal N}=2$ supersymmetry \cite{Witten81, Cooper95} with zero superpotential since there are two supercharges $Q_{1,2}$ that generate the transformation
$
|\varphi\rangle_{odd}=E_\varphi^{-1/2}Q_{1,2}|\varphi\rangle_{even}.
$
Here $|\varphi\rangle_{even}$ and $|\varphi\rangle_{odd}$ are the degenerate eigenstates that satisfy $P|\varphi\rangle_{even}=|\varphi\rangle_{even}$ and $P|\varphi\rangle_{odd}=-|\varphi\rangle_{odd}$, and $E_\varphi$ is the eigenenergy. $\sqrt{H}$ can be obtained by diagonalizing the Hamiltonian in the many-particle space and then take the square root of the diagonal matrix. The explicit form of $Q_{1,2}$ and $\sqrt{H}$ are provided in the appendix A for the case of four Majorana modes.

The degeneracy of states with opposite parities enables the resonant tunneling of a single-electron at zero voltage bias  \cite{Liang14,Tarasinski15, Hsieh16} and thus a conductance peak appear as the signature for the SUSY here. In the following we propose a setup with one dimensional (1D) topological superconductors to realize a supersymmetric closed chain and explore relative novel phonemena.

\emph{Experimental realization.---} 
Because 1D topological superconductors have relatively large minigaps \cite{Alicea12} and candidate materials such as semiconducting nanowires with proximity-induced superconductivity have been fabricated successfully, we adopt two such nanowires to form a closed chain of four coupled Majorana modes. As shown in Fig.~\ref{f1}(b), on a big superconducting ring there are two parallel nanowires (blue) with their ends connected by Josephson junctions (yellow). There is a Zeeman field in the $x$ direction parallel to the nanowires to induce the topological superconductivity and four Majorana modes $\gamma_{1,2,3,4}$ residing at the ends. An applied magnetic flux $\Phi$ in the $z$ direction penetrates through the ring to tune the phase shift across the inter-wire Josephson junctions. This field is much smaller than the field along the nanowire. There is also a reference junction with high impedance and Josephson energy to suppress the phase fluctuation and ensure the phase drop mainly across the inter-wire junctions. 

We first consider a simple but important case that the two wires are identical, but the two junctions can be different. The explicit Hamiltonian is given by $H=H_L+H_R+H_{\Gamma}$
where 
\begin{align}\label{Hj}
\!\!\!H_\beta\!=\!\!\!\int_{0}^{l} \!\!\!\!d x & \psi_{\beta\sigma}^\dag(x)\!\left(\!-\frac{\partial_x^2}{2m^*}\!-\!\mu\!+\!i\alpha \sigma_y \partial_x\!+\!V_x\sigma_x\!\right)_{\sigma\sigma'}\!\!\!\!\!\!\psi_{\beta\sigma'}(x)  \\ \nonumber
 &+\int_{0}^{l} d x[|\Delta| e^{i\theta_\beta}\psi_{\beta\uparrow}^\dagger(x)\psi_{\beta\downarrow}^\dagger(x)+h.c.] 
\end{align}
with $\beta=L,R$, which is the Hamiltonian for each nanowire with length $l$ which combines SOC with strength $\alpha$, Zeeman energy $V_x$ and superconductivity with a gap function $|\Delta| e^{i\theta_\beta}$, and
\begin{eqnarray}\label{Hg}
\!\!\!H_{\Gamma}\!=\!-\!\sum_{\sigma=\uparrow,\downarrow}\!(\Gamma_0\psi_{L\sigma}^\dagger(0) \psi_{R\sigma}(0)\!+\!\Gamma_l\psi_{L\sigma}^\dagger(l) \psi_{R\sigma}(l)\!+\!h.c.)
\end{eqnarray}
describes the the single-electron tunneling across the junctions with strength $\Gamma_{0,l}>0$. The phase shift across the junctions is given by $\theta=\theta_R-\theta_L=2\pi \Phi/\Phi_0$.

To conveniently analyze the couplings between Majorana modes, we adopt the Kitaev's model on 1D spinless p-wave superconductor \cite{Kitaev01} which captures the nature of topological superconductivity in the nanowires. The Hamiltonian is given by $H'=H_{L}'+H_{R}'+H_{\Gamma}'$ where $H_\beta'=\sum_{x=1}^{n-1}\left(-w a_{\beta,x}^\dagger a_{\beta,x+1}+|\Delta_p|e^{i\theta_\beta}a^\dagger_{\beta,x}a^\dagger_{\beta,x+1}+ h. c. \right)$ which describe the left and right spinless p-wave superconductor with $w$ the hopping integral and $|\Delta_p|e^{i\theta_\beta}$ the superconducting gap functions, and $H_{\Gamma}'\!=-\Gamma_0' a_{L1}^\dagger a_{R1}-\Gamma_l' a_{Ln}^\dagger a_{Rn}+h.c.$ with $\Gamma_{0,l}'>0$, which describes the inter-wire single-particle tunneling across the junctions  \cite{Alicea12}. We define $a_{\beta,x}=e^{i\theta_\beta/2}(ib_{\beta,2x-1}+b_{\beta,2x})$ with $b_{\beta,2x-1}$ and $b_{\beta,2x}$ the Majorana operators.

\begin{figure}[t]
\psfig{figure=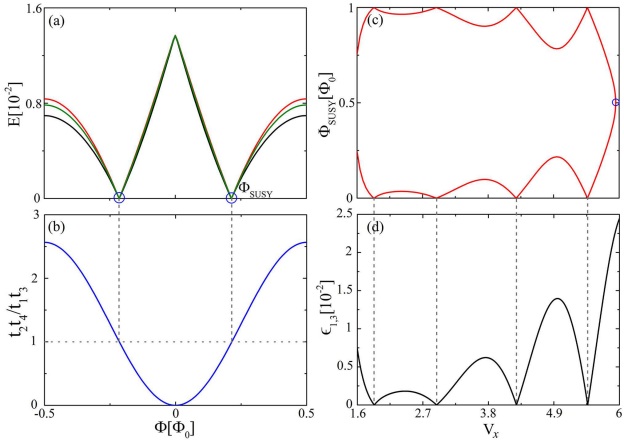,width=8.4cm} \caption{\label{f2} Emergent SUSY tuned by fluxes through the SQUID. (a) Lowest-energy spectra with respect to magnetic fluxes through the SQUID. The red, green and black curves respectively correspond to $\{\Gamma_0, \Gamma_l\}=\{3, 3\},  \{4, 2.25\}$ and $\{5,1.8\}$, which give the same $\Gamma_0\Gamma_l$ and reach zero at the same flux $\Phi_{SUSY}\approx0.213\Phi_0$ where SUSY is obtained (blue circles). The other parameters are $\mu=0, V_x=5, \alpha=1,|\Delta|=1$ and $L=15$ which is discretized into 180 sites. (b) Flux dependence of $t_2t_4/t_1t_3$. $t_1t_3=t_2t_4$ coincides with the appearance of gapless excitations. (c)  Oscillatory dependence of $\Phi_{SUSY}$ on Zeeman energy $V_x$. We use $\{\Gamma_0, \Gamma_l\}=\{4, 2.25\}$ for (b) and (c). (d) Energy splitting due to hybridization of Majorana modes in the same nanowire at different Zeeman energy. 
 }
\end{figure}

We first consider the case $w=|\Delta_p|$ that the Majorana modes stay locally at the edge site, which means that we can write 
$a_{L1} \rightarrow i \frac{1}{2}e^{i\theta_L/2}\gamma_1$, $a_{Ln} \rightarrow \frac{1}{2}e^{i\theta_L/2}\gamma_2$, $a_{R1} \rightarrow i \frac{1}{2}e^{i\theta_R/2}\gamma_4$ and $a_{Rn} \rightarrow \frac{1}{2}e^{i\theta_R/2}\gamma_3$ \cite{Alicea12, Jiang11}, 
leading to $H_\Gamma'=it_2\gamma_2\gamma_3+it_4\gamma_4\gamma_1$ with 
$t_2=\frac{\Gamma_0'}{2}\sin\frac{\theta}{2}, \ \  t_4=-\frac{\Gamma_l'}{2}\sin\frac{\theta}{2}$

which indicates $t_2t_4<0$. When $w\neq|\Delta_p|$, the wavefunctions of Majorana modes exponentially decay from the edges into the bulk, leading to reduced amplitude at the edges. As a consequence, $t_2$ and $t_4$ should be reduced by multiplying a factor $g<1$, but their relative sign does not change. On the other hand, we have the couplings $t_1\gamma_1\gamma_2$ and $t_3\gamma_3\gamma_4$ because the decayed Majorana modes on the same wire inevitably overlap in any realistic wires with finite length. Considering that the two wires are identical, we have $\gamma_4$ identical to $\gamma_1$ and $\gamma_3$ identical to $\gamma_2$ in terms of their locations in Fig.~\ref{f1}(b), and the coupling $t_1\gamma_1\gamma_2$ and $-t_3\gamma_4\gamma_3$ should also be equivalent, leading to $t_1=-t_3$. Since such intra-wire hybridizations correspond to energy splittings $\epsilon_1=|t_1|$ and $\epsilon_3=|t_3|$, we have
\begin{equation}\label{t2t4overt1t3}
\frac{t_2t_4}{t_1t_3}=\frac{g^2\Gamma_0'\Gamma_l'}{4\epsilon_1^2}\sin^2\frac{\pi\Phi}{\Phi_0},
\end{equation}
which indicates that $t_1t_3=t_2t_4$ can be obtained by tuning $\Phi$ when $g^2\Gamma_0'\Gamma_l'/4\epsilon_1^2\ge1$. Accordingly, the zero energy excitations appear at 

\begin{equation}\label{PhiSUSY}
\Phi_{SUSY}=\pm \frac{\Phi_0}{\pi}\arcsin\frac{2\epsilon_1}{g\sqrt{\Gamma_0'\Gamma_l'}}.
\end{equation}
Here we should notice that the change of wavefunctions of the Majorana modes due to these weak couplings are ignorable, which is the reason why we can analyze the couplings separately.

Now we numerically solve the Hamiltonian of nanowires in Eq.~(\ref{Hj}) and (\ref{Hg}) to testify the above analysis. Let us explore the lowest-energy spectra with respect to the magnetic flux. By using the substitutions $x_\alpha=m^*\alpha x$, $E_\alpha=m^*\alpha^2$ to recast the Hamiltonian into a dimensionless form and then solving the corresponding tight-binding Bogoliubov-de Gennes (BdG) equations, we obtain the lowest-energy spectra as shown in Fig.~\ref{f2}(a). The three curves correspond to three groups of parameters which have different $\Gamma_{0,l}$ but the same $\Gamma_0\Gamma_l$. At the magnetic flux around $\Phi_{SUSY}\approx\pm0.213\Phi_0$ all three curves reach zero, which indicates the emergence of SUSY. We get unchanged $\Phi_{SUSY}$ when keeping $\Gamma_0\Gamma_l$ to be constant. This property is reflected in Eq.~(\ref{PhiSUSY}) in the form that $\Gamma_0'\Gamma_l'$ is the characteristic value not the $\Gamma_0'$ and $\Gamma_l'$ separately. If we consider an additional small amount of flux threading the space between two nanowires, which is a situation in real experiments, $\Phi_{SUSY}$ is shifted a little to recover the SUSY (see appendix C).

Now we numerically obtain $t_2t_4/t_1t_3$ to check the correspondence between $t_2t_4=t_1t_3$ and the appearance of SUSY. We first consider a single nanowire where only the coupling $it_1\gamma_1\gamma_2$ or $it_3\gamma_3\gamma_4$ is available. By solving $H_L$ with the parameters given in Fig.~\ref{f2}(a), we obtain $|t_1|=\epsilon_1\approx0.0139$. We have $|t_3|=|t_1|$ because two nanowires are the same. The situation with only $it_2\gamma_2\gamma_3$ can be found in the setup with two long nanowires $(t_1=t_3\approx 0)$ and $\Gamma_0=0$, and then we obtain $|t_2|=\epsilon_2\approx 0.0297|\sin(\pi\Phi/\Phi_0)|$ for $\Gamma_l=2.25$, where $E_2$ is the first finite-energy excitation. Similarly we get $|t_4|=\epsilon_4\approx 0.0167|\sin(\pi\Phi/\Phi_0)|$ for $\Gamma_0=4$. Considering the same sign of $t_1t_3$ and $t_2t_4$, we obtain $t_2t_4/t_1t_3\approx2.57\sin^2\pi\Phi/\Phi_0$ which is consistent with Eq.~(\ref{t2t4overt1t3}). By drawing this relation in Fig.~\ref{f2}(b), we can observe an exact correspondence between $t_3t_4=t_1t_2$ and the appearance of gapless excitations by comparing Fig.~\ref{f2}(a) and~(b), which proves that our set-up can realize a closed chain of Majorana modes with nearest neighbour coupling where the SUSY can be obtained.

\emph{Oscillation of $\Phi_{SUSY}$ as signature of Majorana modes.---}Let us study the dependence of $\Phi_{SUSY}$ on the Zeeman energy $V_x$. Since the Majorana wavefunctions depend on $V_x$, so do $t_j$ and $t_2t_4/t_1t_3$ as well. As a consequence, when we change $V_x$ after obtaining $t_1t_3=t_2t_4$, this equality should be rebuilt by finding a new $\Phi_{SUSY}$ in general.  For a typical group of parameters, we get oscillatory curves for $\Phi_{SUSY}(V_x)$ as shown in Fig. \ref{f2}(c), which has three noteworthy features. First of all, the curves oscillate in a similar way to $\epsilon_1(V_x)$ shown in Fig.~\ref{f2}(d), indicating $t_1$ and $t_3$ as the dominant role in changing $\Phi_{SUSY}$. Moreover, the curves repeat in every regime of $\Phi\in[m,m+1]\Phi_0$ with $m$ an integer and  are symmetric with respect to the axises $\Phi=m\Phi_0,m\Phi_0/2$ because $t_2t_4/t_1t_3\propto\sin^2(\pi\Phi/\Phi_0)$ is an even function of $\Phi$ with a period of $\Phi_0$. Here Fig.~\ref{f2}(c) is for the regime with $m=0$. Last but not least, with increasing $V_x$ the lower and upper curves reach at $\Phi_{SUSY}=0.5\Phi_0$ as noted by the blue circle in Fig.~\ref{f2}(c) and then $\Phi_{SUSY}$ do not exist within a range of larger $V_x$ where the increased $\epsilon_1$ makes $2\epsilon_1/g\sqrt{\Gamma_0'\Gamma_l'}>1$ in Eq.~(\ref{PhiSUSY}).

To our knowledge, this phenomenon that fluxes realizing zero-bias conductance peak oscillate with $V_x$ with above three features have not been reported in any other systems, thus serving as a unique signature to test the existence of Majorana mode. To emphasize, the conductance peak is not blurred by extra Cooper-pair tunnelling through the junctions, showing its advantage over the fractional Josephson effect on detecting Majorana modes. Moreover, since the current-phase relation is not explored here, the parity conservation is not required for the observation of $\Phi_{SUSY}$. Since the oscillation of zero-energy splitting with the Zeeman fields has been observed experimentally in a $0.9\mu m$ $\textrm{InAs}$ nanowire with an epitaxial aluminium shell \cite{Albrecht16}, the same nanowires can be adopted for our proposal and the corresponding oscillation of $\Phi_{SUSY}$ should be observed if that splitting is caused by hybridization of Majorana modes.

\begin{figure}[t]
\psfig{figure=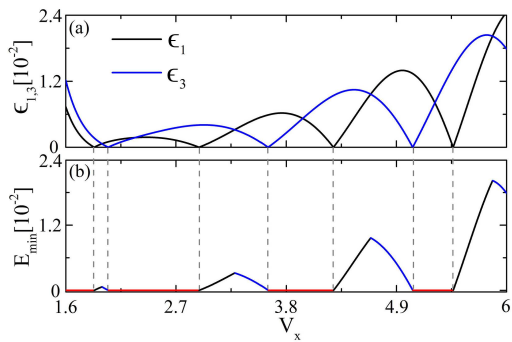,width=6.8cm} \caption{\label{f3} 
 (a) Oscillatory dependence of energy splitting on Zeeman energy in different nanowires. (b) Smallest value of lowest-energy excitations within $\Phi\in[0,\Phi_0]$ at different $V_x$. Red lines show the regime where gapless excitations are available. We use $\{\Gamma_0, \Gamma_l\}=\{3, 3\}$ and other parameters are the same as Fig.~\ref{f2}.}
\end{figure}

So far, we have focused on the setup with two same nanowires. Now we study the case with different nanowires by increasing the strength of spin-orbital coupling of the right nanowire by $10\%$. We find different oscillation curves of energy splitting compared with the unchanged left nanowire, as shown in Fig.~\ref{f3}(a). In particular, the two curves touch zero at different $V_x$. Because touching zero indicates a sign change of the corresponding $t_1$ or $t_3$ \cite{Sarma12,Cheng09,Mizushima10,Cheng10}, the sign of $t_1t_3$ oscillates as well with the Zeeman energy. On the other hand, the sign of  $t_2t_4$ is fixed, which means that $t_1t_3$ and $t_2t_4$ have opposite signs in some regimes of $V_x$ where SUSY cannot be obtained. To testify this, we numerically study the smallest value of the lowest-energy excitations within $\Phi\in[0,\Phi_0]$ at different $V_x$ and the energy spectra are given in Fig.~\ref{f3}(b). The gapless excitations are available in separated regimes with boundaries where the sign of $t_1t_3$ reverses. For $V_x$ outside these regimes, all excitations are gapful. In particular for the cases with $\Gamma_0=\Gamma_l$, we can prove that the smallest values are obtained at $t_2=t_4=0$, i.e. $\Phi=0$, and the value is the smaller one between $|t_1|$ and $|t_3|$, as shown in appendix B.

\emph{Summaries and Discussions.---} In this letter, we have proved a supersymmetry in a closed chain of nearest-neighbour coupled Majorana modes by tuning only one arbitrary coupling. We have adopted two nanowires with ends connected by Josephson junctions as a setup for experimental realization of a closed chain of four coupled Majorana modes. By using a magnetic flux $\Phi$ to tune the Josephson couplings, we have obtained the supersymmetry at $\Phi= m\Phi_0\pm \Phi_{SUSY}$ which is signaled by a zero-bias conductance peak. In particular, $\Phi_{SUSY}$ has an oscillatory dependence on the Zeeman field parallel to the nanowires, which is a unique phenomena and clear evidence for the existence of Majorana modes.

Oscillation of zero-energy splitting and fractional Josephson effect are two nontrivial phenomena of Majorana modes. Due to the complexity of real experiments, mechanisms other than Majorana modes may also realize either phenomenon, but their chances to realize both phenomena together should be much less. Therefore, the oscillatory $\Phi_{SUSY}$, which is based on the interplay of the two phenomena, is a more convincing signature for the existence of Majorana modes than the two phenomena working separately. Our system thus has a large potential to help facilitate notable progress in the experimental study of topological superconductivity.

Apart from the setup shown in Fig.~\ref{f1}(b), there are other possible methods to realize our proposal with cutting-edge techniques. Recently a wire-like thin layer Al has been produced lithographically on a 2D layer of electron gas in order to fabricate a 1D topological superconductor \cite{Suominen17}. The same technique can be adopted to fabricate two parallel 1D topological superconductors with ends connected by deposited insulating barriers. Another method is to apply a gate voltage along the centerline of the nanowire to push the electron gas to the right and left surface, which effectively ``cut'' one nanowire into two parallel 1D electron gas \cite{DengPrivate}, thus achieving four Majorana modes on a single wire. Moreover, four nearest-neighbour coupled Majorana modes are realized as natural situations for the second-order topological superconductors, which is a superconducting generalization of square second-order topological insulators with four corner states \cite{Song17, Schindler17,Langbehn17}.

\vspace{3mm}
\noindent {\it Acknowledgements.---} The authors are grateful for Ippei Danshita, Mingtang Deng, Timothy H. Hsieh, Takuto Kawakami, Laurens W. Molenkamp,  Simon Trebst, Zhi Wang and Hongqi Xu for helpful discussions. This work is supported by the MEXT-Supported Program for the Strategic Research Foundation at Private Universities ``Topological Science'' (No. S1511006). The work of M. N. is supported in part by the JPSJ Grant-in-Aid for Scientific Research (KAKENHI No. 16H03984) and ``Topological Materials Science'' (KAKENHI No. 15H05855) from the MEXT of Japan.

\appendix

\section{Supercharges for a supersymmetric closed chain of four interacting Majorana modes}

In this appendix, we present the explicit form of the supercharges in a supersymmetric closed chain of four Majorana modes with nearest-neighbour couplings. The corresponding Hamiltonian is given by 
\begin{equation}
H_{eff}=i(t_1\gamma_1\gamma_2+t_2\gamma_2\gamma_3+t_3\gamma_3\gamma_4+t_4\gamma_4\gamma_1).
\end{equation}
By defining two Dirac operators $c_1=\gamma_1+i\gamma_2$ and $c_2=\gamma_3+\gamma_4$, we have the many-particle basis $\{|00\rangle, |11\rangle, |10\rangle, |01\rangle\}$ with $|00\rangle=|\phi_0\rangle, |11\rangle=c_1^\dagger c_2^\dagger|\phi_0\rangle, |10\rangle=c_1^\dagger|\phi_0\rangle$ and $|01\rangle=c_2^\dagger |\phi_0\rangle$. Under this basis, the matrix form of $H_{eff}$ is given by 
\begin{equation}
\hat{H}_{eff}=
\begin{pmatrix}
-t_1-t_3 &t_4-t_2 &0 &0 \\
t_4-t_2 &t_1+t_3 &0 &0 \\
0 &0 &t_1-t_3 &-t_2-t_4 \\
0 &0 &-t_2-t_4 &t_3-t_1
\end{pmatrix}.
\end{equation}
By imposing the condition for SUSY $t_1t_3=t_2t_4$, we get the energy levels $E=\pm \epsilon$ with $\epsilon=\sqrt{t_1^2+t_2^2+t_3^2+t_4^2}$ which are two-fold degenerate. In order to construct the supercharges, all energy levels need to be non-negative, so we shift the Hamiltonian by a positive constant $h\epsilon$ with $h\ge 1$. We write the shifted Hamiltonian as $H_{SUSY}=H_{eff}+h\epsilon$ and the energy levels are $E_1=(h-1)\epsilon$ and $E_2=(h+1)\epsilon$. The degenerate states at $E_1$ are obtained as
\begin{equation}\label{excitations}
\begin{split}
|\varphi\rangle_{even}=\frac{1}{A_1}[(t_2-t_4)|00\rangle+(\epsilon-t_1-t_3)|11\rangle,\\
|\varphi\rangle_{odd}=\frac{1}{B_1}[(t_4+t_2)|10\rangle+(\epsilon+t_1-t_3)|01\rangle,
\end{split}
\end{equation}
and the degenerate states at $E_2$ are obtained as
\begin{equation}\label{excitations}
\begin{split}
|\varphi\rangle_{even}=\frac{1}{A_2}[(t_2-t_4)|00\rangle-(\epsilon+t_1+t_3)|11\rangle,\\
|\varphi\rangle_{odd}=\frac{1}{B_2}[(t_4+t_2)|10\rangle+(-\epsilon+t_1-t_3)|01\rangle,
\end{split}
\end{equation}
where $A_1,B_1,A_2$ and $B_2$ are coefficients for normalization. We thus find the two-fold degeneracy of states with opposite fermion parities. 

According to the analysis in the main text, there are two fermionic operators
\begin{eqnarray}
Q_1=\gamma'\sqrt{H_{SUSY}}, \ \ \ Q_2=\gamma''\sqrt{H_{SUSY}},
\end{eqnarray}
which satisfy the algebra
\begin{eqnarray}
\{P,Q_i\}=0,\ \ \{Q_i,Q_j\}=2\delta_{ij}H_{SUSY}
\end{eqnarray}
for $i,j\in\{1,2\}$. This indicate an ${\cal N}=2$ supersymmetry and $Q_{1,2}$ are the two supercharges. For the case of four Majorana modes here, we obtain
\begin{equation}
\gamma'=\frac{t_2}{\sqrt{t_1^2+t_2^2}}\left(\gamma_1+\frac{t_1}{t_2}\gamma_3\right),\ \ \ \gamma''=\frac{t_3}{\sqrt{t_2^2+t_3^2}}\left(\gamma_2+\frac{t_2}{t_3}\gamma_4\right)
\end{equation}
and
\begin{eqnarray}
\sqrt{H_{SUSY}}=\frac{1}{A\sqrt{\epsilon}}H_{eff}+B\sqrt{\epsilon},
\end{eqnarray} 
with
\begin{eqnarray}
A=\sqrt{h+1}+\sqrt{h-1},\ \ \ \ B=\frac{\sqrt{h+1}+\sqrt{h-1}}{2}.
\end{eqnarray} 
We can easily check that $(H_{eff}/A\sqrt{\epsilon}+B\sqrt{\epsilon})^2=H_{eff}+h\epsilon$.

\section{Quasiparticle excitations for non-supersymmetric closed chains of four interaction Majorana modes}

In this appendix, we study the quasiparticle excitations in a closed chain of four interacting Majorana modes with $t_2t_4/t_1t_3<0$, which is a situation in the regimes of $V_x$ in Fig.~3(b) where only finite energy excitations are available. Here we rewrite the $H_{eff}$ in terms of Dirac operators as
\begin{align}
H_{eff}=&t_1(c_1^\dagger c_1-c_1c_1^\dagger)-(t_2+t_4)c_2^\dagger c_1-(t_2+t_4)c_1^\dagger c_2 \\ \nonumber
+&(t_2-t_4)c_1c_2+(t_4-t_2)c_1^\dagger c_2^\dagger+t_3(c_2^\dagger c_2-c_2 c_2^\dagger),
\end{align}
which has the matrix form 
\begin{equation}
\hat{H}_{BdG}=\frac{1}{2}
\begin{pmatrix}
2t_1 &-(t_2+t_4) &0 &t_4-t_2 \\
-(t_2+t_4) &2t_3 &t_2-t_4 &0 \\
0 &t_2-t_4 &-2t_1 &t_2+t_4 \\
t_4-t_2 &0 &t_2+t_4 &-2t_3
\end{pmatrix}
\end{equation}
under the BdG basis $\{c_1^\dagger, c_2^\dagger, c_1,c_2\}$. By diagonalizing the matrix, we get four quasiparticle excitations with energy 
\begin{equation}\label{BdGexcitation}
\epsilon=\pm\frac{1}{\sqrt{2}}\sqrt{\sum_{j=1}^4t_j^2\pm\sqrt{\left(\sum_{j=1}^4t_j^2\right)^2-4(t_1t_3-t_2t_4)^2}},
\end{equation}
where the positive and negative excitations are symmetric due to particle-hole symmetry. Because $t_1t_3\neq t_2t_4$ due to $t_2t_4/t_1t_3<0$, there is no gapless excitations according to Eq.~(\ref{BdGexcitation}). Therefore, no matter how we change the magnetic flux $\Phi$ through the ring in Fig.~1(b) of the main context to tune $t_{2,4}$, we cannot obtain SUSY. 

Now we focus on a special case with $|t_2|=|t_4|$, where we can prove that the smallest values of the lowest excitations appear at $t_2=t_4=0$, i.e. $\Phi=0$. Without loss of generality, we consider $t_2=t_4$ and $t_1t_3<0$, then we have the first excitation as

\begin{equation}\label{1st}
\epsilon_1=\frac{1}{\sqrt{2}}\sqrt{(t_1^2+t_3^2+2t_2^2)-\sqrt{(t_1^2-t_3^2)^2+4t_2^2(t_1+t_3)^2}}.
\end{equation}
It is straightforward to prove $\epsilon_{1}\ge \epsilon_{1}(t_2=t_4=0)$ for any $t_{1,3}$, thus giving the smallest value 
\begin{equation}\label{Emin}
E_{min}=|\epsilon_1(t_2=t_4=0)|=\left|\frac{t_1+t_3}{2}+\frac{|t_1-t_3|}{2}\right|,
\end{equation}
which is the smaller one between $|t_1|$ and $|t_3|$ as shown in Fig.~3(b) in the main text.

\section{Effects of additional fluxes threading the closed chain of Majorana modes}

In real experiments, we consider an external ring much larger than the closed chain of two nanowires and two junctions. Nevertheless, when fluxes penetrate the external ring, it may not be avoided for a small amount of flux to thread the closed chain. Here we show that the system can still be tuned to the supersymmetric state in presence of such additional fluxes. 

\begin{figure}[t]
\psfig{figure=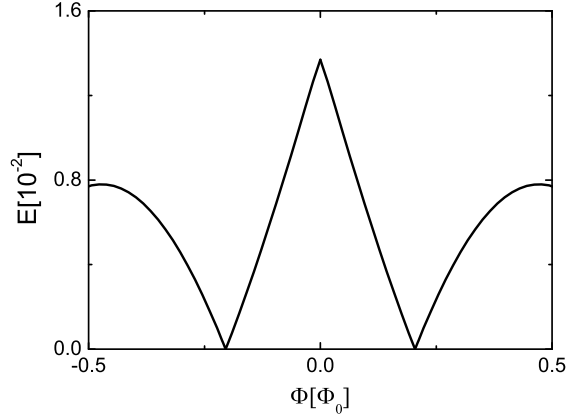,width=8.5cm} \caption{\label{fsupp} 
Lowest-energy spectrum with respect to magnetic fluxes through the loop formed by external superconducting ring and junction between $\gamma_2$ and $\gamma_3$ in Fig. 1(b) of the main text. The additional flux threading the closed chain of coupled Majorana modes is $\Phi'=0.1\Phi$. Other parameters are the same as the green curve of Fig. 2(a) in the main text. }
\end{figure}

We analyze with Kitaev chains as the same with the main text, where the tunneling term related to Josephson effect is given by
\begin{equation}
H_{\Gamma}'\!=-\Gamma_1 a_{L1}^\dagger a_{R1}-\Gamma_n a_{Ln}^\dagger a_{Rn}+h.c..
\end{equation} In the topological regime,  we have $a_{L1} \rightarrow i \frac{1}{2}ge^{i\theta_{L1}/2}\gamma_1$, $a_{Ln} \rightarrow \frac{1}{2}ge^{i\theta_{Ln}/2}\gamma_2$, $a_{R1} \rightarrow i \frac{1}{2}ge^{i\theta_{R1}/2}\gamma_4$ and $a_{Rn} \rightarrow \frac{1}{2}ge^{i\theta_{Rn}/2}\gamma_3$. The phase shift across a junction is determined by the fluxes surrounded by the junction and external ring, and thus we have 
\begin{equation}\label{t2t4overt1t3}
\theta_{R1}-\theta_{L1}=2\pi\frac{\Phi+\Phi'}{\Phi_0}, \ \ \ \theta_{Rn}-\theta_{Ln}=2\pi\frac{\Phi}{\Phi_0},
\end{equation}
where $\Phi'$ is the additional flux threading the closed chain of coupled Majorana modes. We thus obtain 
\begin{equation}
H_\Gamma'=it_2\gamma_2\gamma_3+it_4\gamma_4\gamma_1
\end{equation} 
with 
\begin{equation}
t_2=\frac{\Gamma_1}{2}g\sin\frac{\pi\Phi}{\Phi_0}, \ \ \ 
t_4=-\frac{\Gamma_n}{2}g\sin\frac{\pi(\Phi+\Phi')}{\Phi_0}.
\end{equation}
By using that $t_1t_3=-E_1^2$ in the main text, we obtain
\begin{equation}\label{t2t4overt1t3}
\frac{t_2t_4}{t_1t_3}=\frac{g^2\Gamma_1\Gamma_n}{4E_1^2}\sin\frac{\pi\Phi}{\Phi_0}\sin\frac{\pi(\Phi+\Phi')}{\Phi_0}.
\end{equation}
Since the external ring is much larger than the closed chain, we consider $\Phi'\ll \Phi$, in which case SUSY can still be obtained when $g^2\Gamma_1\Gamma_n/4E_1^2\ge 1$, but $\Phi_{SUSY}$ is shifted a little from the value corresponding to $\Phi'=0$. For an example $\Phi'=0.1\Phi_0$, $\Phi_{SUSY}$ changes from $0.213\Phi_0$ in Fig.~2(a) of the main text to $0.205\Phi_0$ in Fig. 4.

\bibliography{reference}

\end{document}